\documentclass[fleqn,twoside]{article}
\usepackage{espcrc2}

\usepackage{graphicx}
\usepackage[figuresright]{rotating}

\newcommand{\AmS}{{\protect\the\textfont2
  A\kern-.1667em\lower.5ex\hbox{M}\kern-.125emS}}

\title{Models for neutrino mass with discrete symmetries}

\author{S.\,Morisi\address[MCSD]{AHEP Group, Institut de F\'{\i}sica Corpuscular --
  C.S.I.C./Universitat de Val{\`e}ncia \\
  Edificio Institutos de Paterna, Apt 22085, E--46071 Valencia, Spain}%
}
       
\begin{document}

\begin{abstract}
Discrete non-abelian flavor symmetries give in a natural way tri-bimaximal (TBM) mixing
as showed in a prototype model. However neutrino mass matrix pattern may be
very different from the tri-bimaximal one if small deviations of TBM will be observed.
We give the result of a model independent analysis for TBM neutrino mass pattern.
\end{abstract}

\maketitle

\begin{figure}[h!]
\begin{center} 
\includegraphics[width=4.cm]{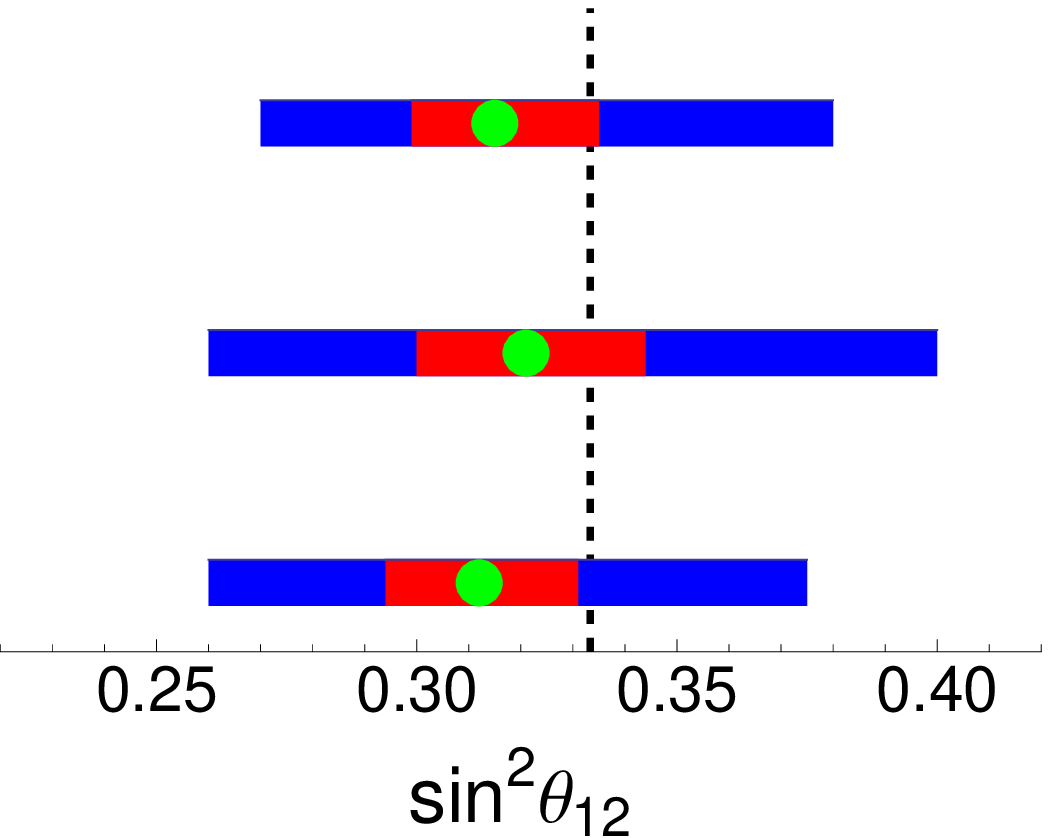}
\vskip2mm
\includegraphics[width=4.cm]{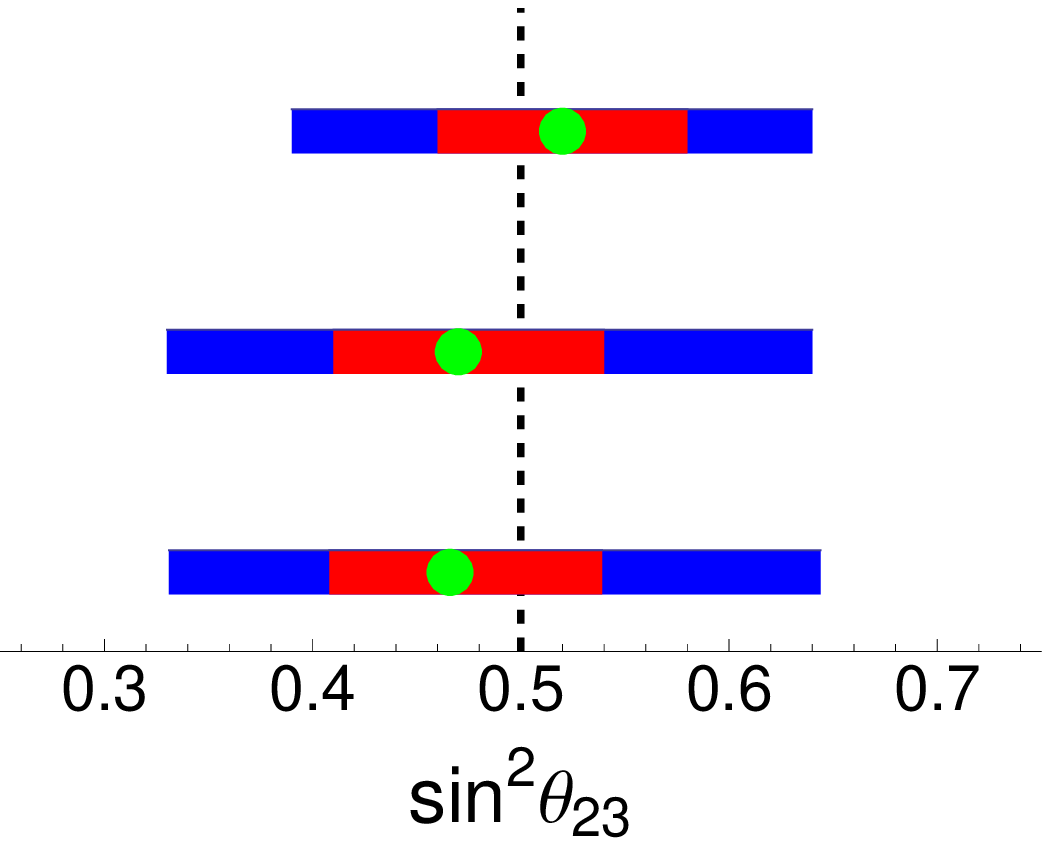}
\vskip2mm
\includegraphics[width=4.cm]{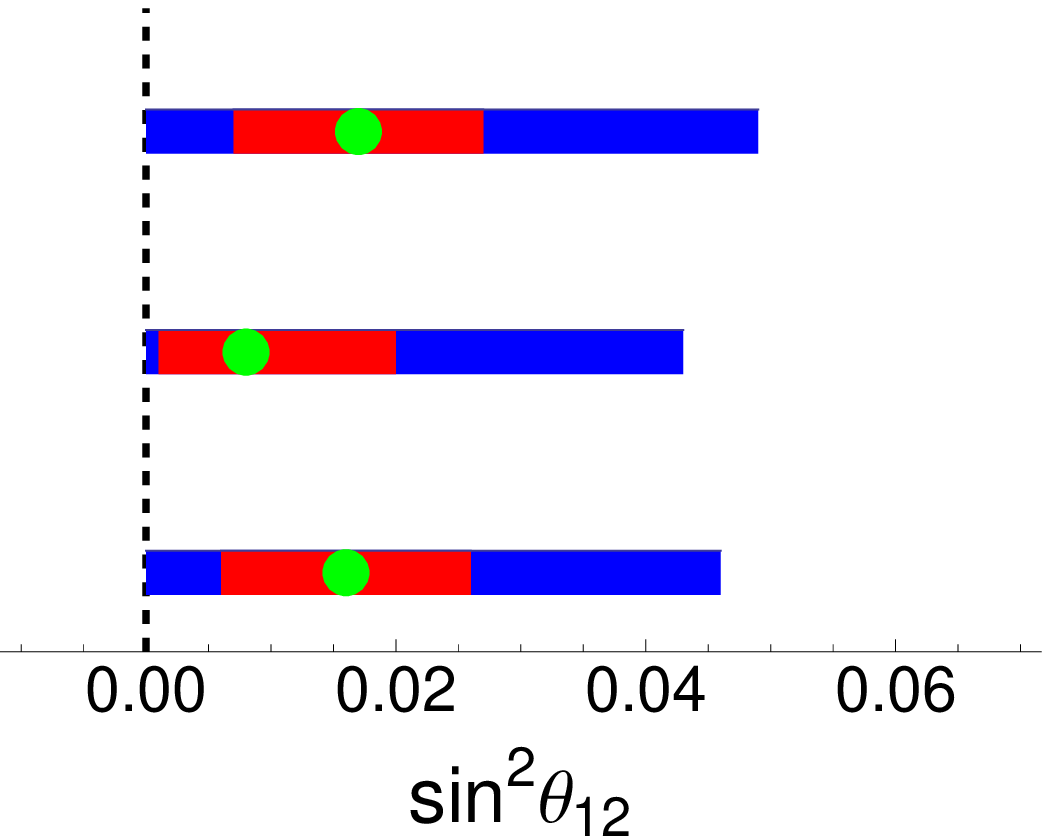}
\caption{The blue and red ranges are respectively the $3\sigma$ and $1\sigma$
values for $\sin^2\theta_{12}$ (up plot), $\sin^2\theta_{23}$ (middle plot) and $\sin^2\theta_{13}$ (bottom plot)
from the Ref.\cite{Schwetz:2008er} (up band), \cite{GonzalezGarcia:2010er} (middle band) and \cite{Fogli:2005cq} (down band) and the green
points are the best fit point. The vertical line are the TBM values and the $\lambda_C$, $\lambda_C^2$ deviations.}
\label{fig1}
\end{center}
\end{figure}

Neutrino data %\cite{Schwetz:2008er,GonzalezGarcia:2010er,Fogli:2005cq} 
are in well agreement (approximatively within $1\sigma$)
with the so called tri-bimaximal mixing \cite{Harrison:2002er}, giving maximal
atmospheric angle $\sin^2\theta_{23}=1/2$, zero reactor angle  $\sin^2\theta_{13}=0$
and trimaximal solar angle  $\sin^2\theta_{12}=1/3$.
This is shown in fig.\ref{fig1} where we give the result of the fits  of three different groups 
\cite{Schwetz:2008er,GonzalezGarcia:2010er,Fogli:2005cq} represented by three 
horizontal bands where the blue band is the $3\sigma$ region,  the red band is the $1\sigma$ region, 
and the green point is the best fit value.  

Non-abelian discrete symmetries has been extensively used in order to explain TBM mixing for neutrino, see for 
instance \cite{Altarelli:2010gt} and reference therein.
We provide a simple example %that reproduce tribimaximal mixing showing a prototype model
based on the discrete abelian flavor symmetry $A_4$\cite{Babu:2002dz,Altarelli:2005yp,Altarelli:2005yx}, namely the even permutations
of four objects. This is the smallest finite group with triplet irreducible representation
and seems very adequate to accommodate the three flavor families of the Standard Model.

We consider the Altarelli-Feruglio model defined in table\,\ref{tab1}.
\begin{table}[h!]
\begin{center}
\begin{tabular}{lcccccccc}
\hline
&$L$&$e^c$&$\mu^c$&$\tau^c$ &$h_{u,d}$&$\varphi_T$&$\varphi_S$&$\xi$\\
\hline
$SU(2)$ &2 &1 &1 &1 &2 &1&1&1\\
$A_4$&3&1&$1'$&$1''$&$1$&$3$&$3$&$1$\\
$Z_3^{\mbox{\tiny aux}}$&$\omega$ &$\omega^2$ &$\omega^2$ &$\omega^2$ &1 &1 &$\omega$&$\omega$\\
\hline
&&&&&&&&\\
\end{tabular}
\caption{Matter assignment of a prototype model for TBM mixing.}\label{tab1}
\end{center}
\end{table}
The flavon field $\varphi_T$
is coupled only to the charged fermions $Ll^ch\varphi_T$  and $\varphi_S$
is coupled only to the dimension five operator $LLhh\varphi_S$ in order to preserve
the additional $Z_3^{\mbox{\tiny aux}}$. 
After that the flavon fields take vev, $A_4$ is spontaneously broken as below
\begin{equation}\label{eq1}
\begin{array}{lll}
\langle \varphi_T \rangle \sim (1,0,0)\,&: & A_4\to Z_3\,;\\
\langle \varphi_S \rangle \sim (1,1,1)\,&: & A_4\to Z_2\,.
\end{array}
\end{equation}
So $A_4$ is spontaneously broken into $Z_3$ in the charged fermion sector 
while into $Z_2$ in the neutrino sector where $Z_2$ and $Z_3$ are two subgroups of
$A_4$. The misalignment between the two sectors is at the origin of large lepton mixing. 
In general this is possible thanks to auxiliary abelian symmetries 
that distinguish the charged and neutral lepton sectors.
In the prototype model we are presenting such a auxiliary symmetry is $Z_3^{\mbox{\tiny aux}}$.
  
It is well know that the alignments $(1,0,0)$ or $(1,1,1)$ in eq.\,(\ref{eq1}) are natural in $A_4$
since they leave unbroken the $T$ and the $S$ generators of $A_4$ respectively. 
However assuming complex vevs different solutions can be found from the minimization
of the potential, see \cite{Lavoura:2007dw,Morisi:2009sc,Toorop:2010ex}. In general the solution of the potential
containing both the scalar fields $\varphi_S$ and $\varphi_T$ is not given by  eq.\,(\ref{eq1})
unless terms mixing $\varphi_S$ and $\varphi_T$ like $|\varphi_S|^2|\varphi_T|^2$ are neglected 
(this is not possible by means of symmetries). 
It is possible to solve such a problem assuming SUSY\,\cite{Altarelli:2005yp} or extra-dimension\,\cite{Altarelli:2005yx}. 

As a result of the model in table\,(\ref{tab1}) the neutrino mass matrix is $\mu-\tau$ invariant and trimaximal,
that is $m_{\nu_{11}}+m_{\nu_{12}}=m_{\nu_{22}}+m_{\nu_{23}}$ as below
\begin{equation}\label{TBMp}
m_\nu\equiv\left(
\begin{array}{ccc}
x&y&y\\
y&x+z&y-z\\
y&y-z&x+z\\
\end{array}
\right).
\end{equation}
Shortly the neutrino mass matrix has TBM texture.
In general, in almost of the models studied in literature,
 the three angles receive corrections of the same order from next to leading order contributions, 
that is $\sin^2\theta_{23}=1/2+\mathcal{O}(\epsilon)$, $\sin\theta_{13}=\mathcal{O}(\epsilon)$ and 
$\sin^2\theta_{12}=1/3+\mathcal{O}(\epsilon)$ where $\epsilon$ is a small perturbation. 
In order to satisfy solar neutrino data, $\epsilon$ can be almost of order $\lambda_C^2$,
and the deviation of  the reactor angle from TBM must be of the same order. If a larger value
for the reactor angle will be measured, currently at $3\sigma$ the reactor angle can be as larger as
$\lambda_C$, most of the models will be ruled out.

\begin{figure}[h!]
\begin{center} 
\includegraphics[width=4.4cm]{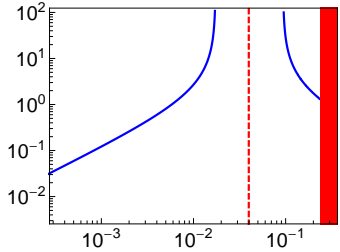}
\caption{$\Delta_e$ vs $\sin\theta_{13}$, the vertical line is the $\lambda_C^2$ value.}
\label{fig2}
\end{center}
\end{figure}

For small deviation of neutrino mass matrix from TBM pattern, see eq.\,(\ref{TBMp}), we expect
small deviation of TBM mixing.
However assuming small deviations (of order $\lambda_C^2$) of the lepton mixing matrix 
from TBM, the deviations of the neutrino mass matrix from the TBM texture\,(\ref{TBMp}) can be very large
as indicated in \cite{Abbas:2010jw}. In order to parameterize the deviation of the neutrino
mass matrix from the $\mu-\tau$ exchange symmetry and from the trimaximality, that is 
$m_{\nu_{11}}+m_{\nu_{12}}=m_{\nu_{22}}+m_{\nu_{23}}$, in \cite{Abbas:2010jw}
the following parameters has been introduced
\begin{eqnarray}
\Delta_{e}&=&\frac{m_\nu^{e\mu} - m_\nu^{e\tau}}{m_\nu^{e\mu}},~\\
\Delta_{\mu\tau}&=&\frac{m_\nu^{\mu\mu} - m_\nu^{\tau\tau}}{m_\nu^{\tau\tau}}~,~~\\
\Delta_\Sigma&=&\frac{\Sigma_L - \Sigma_R}{\Sigma_R}, 
\end{eqnarray}
where $\Sigma_L = m_\nu^{ee}+(m_\nu^{e\mu} + m_\nu^{e\tau})/2$
and $\Sigma_R = m_\nu^{\mu\tau} + (m_\nu^{\mu\tau} +
m_\nu^{\tau\tau})/2$. 
The result of the model independent analisys is shown
in fig.\,(\ref{fig2}) where a correlation between the parameter $\Delta_e$
and $\sin\theta_{13}$ has been given. Such analysis shows that the deviation of the neutrino mass matrix
from TBM texture is small, namely of order $10\%$,  if $\sin\theta_{13}< 10^{-3}$
that is very small. When $\sin\theta_{13}\sim\lambda_C^2$, in the sensitivity of future experiments,
 the deviation of TBM pattern can 
be very large $\Delta_3 > 1$, see fig.\,(\ref{fig2}).

\begin{figure}[h!]
\begin{center} 
\includegraphics[width=3.8cm]{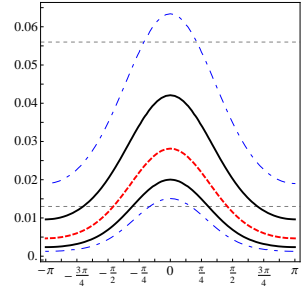}
\caption{$\sin\theta^2_{13}$ vs the Dirac CP phase $\delta$. The central line is for the best
fit, while the other lines are the one and three $\sigma$.}
\label{fig3}
\end{center}
\end{figure}

Also if the model independent analysis %very accurate neutrino data 
do not strictly indicate TBM neutrino  pattern, non-abelian discrete symmetries  
remain a simple and economical description of data. 
For instance, a model with only $A_4$ symmetry (no auxiliary symmetries) and without flavons has been proposed 
in \cite{Morisi:2009sc} where the standard model has been extended assuming extra Higgs
doublets transforming as a triplet of $A_4$. Another example of model based on discrete symmetries
with interesting phenomenological consequence is given in \cite{Meloni:2010aw} with approximate 
Fritzsch texture for light neutrino mass matrix
\begin{equation}
m_\nu\sim\left(
\begin{array}{ccc}
0 &b&0 \\
 b&a&c\\
0&c&d\\
\end{array}
\right)\,.
\end{equation}
The model is based on $S_3$ flavor symmetry, the permutation of three objects with singlets and doublet 
irreducible representations.   
The charged lepton mass matrix is close to diagonal. The two zeros of the  Fritzsch texture,
give a strong correlation between the reactor angle $\theta_{13}$ and the Dirac CP phase $\delta$.
In fig.\,\ref{fig3} we show  such a  correlation. The central line is obtained assuming all the observables at the best fit, and
we have for maximal CP violation $\delta=\pm \pi/2$ that $\sin\theta^2_{13}\approx 0.01$ close to the indication 
of ref.\,\cite{Fogli:2008jx}.

\section*{Acknowledgments}
This work was supported by the Spanish MICINN under grants
FPA2008-00319/FPA and MULTIDARK Consolider CSD2009-00064, by
Prometeo/2009/091, by the EU grant UNILHC PITN-GA-2009-237920 and Juan de la Cierva contract.

\end{document}